\documentclass[11pt]{article}
\usepackage{amssymb}

\usepackage{graphicx}
\usepackage{amsmath}
\usepackage{makeidx}
\usepackage{indentfirst}

\newcounter{resultnum}[section]

\newcounter{conclusionnum}[section]

\newcounter{conditionnum}[section]

\newcounter{conjecturenum}[section]

\newcounter{examplenum}[section]

\newcounter{exercisenum}[section]

\newcounter{lemmanum}[section]

\newcounter{notationnum}[section]
\newtheorem{theorem}{Theorem}[section]

\newcounter{theoremnum}[section]
\newtheorem{definition}{Definition}[section]

\newcounter{definitionnum}[section]

\newcounter{corollarynum}[section]
\newtheorem{remark}{Remark}[section]

\newcounter{remarknum}[section]

\newcounter{propositionnum}[section]

\newcounter{acknowledgementnum}[section]

\newcounter{algorithmnum}[section]

\newcounter{axiomnum}[section]

\newcounter{casenum}[section]

\newcounter{claimnum}[section]

\newcounter{summarynum}[section]

\newcounter{problemnum}[section]

\begin{document}

\title{Fractional Analogous Models in \\
Mechanics and Gravity Theories}
\date{July 29, 2010}
\author{\textbf{Dumitru Baleanu}\thanks{%
On leave of absence from Institute of Space Sciences, P. O. Box, MG-23, R
76900, Magurele--Bucharest, Romania, \newline
E--mails: dumitru@cancaya.edu.tr, baleanu@venus.nipne.ro} \\
\textsl{\small Department of Mathematics and Computer Sciences,} \\
\textsl{\small \c Cankaya University, 06530, Ankara, Turkey } \\
\and 
\textbf{Sergiu I. Vacaru} \thanks{%
sergiu.vacaru@uaic.ro, Sergiu.Vacaru@gmail.com \newline
http://www.scribd.com/people/view/1455460-sergiu} \and \textsl{\small %
Science Department, University "Al. I. Cuza" Ia\c si, } \\
\textsl{\small 54, Lascar Catargi street, Ia\c si, Romania, 700107 } }
\maketitle

\begin{abstract}
We briefly review our recent results on the geometry of nonholonomic
manifolds and Lagrange--Finsler spaces and fractional calculus with Caputo
derivatives. Such constructions are used for elaborating analogous models of
fractional gravity and fractional Lagrange mechanics. \vskip0.2cm


\vskip3pt


\end{abstract}

\tableofcontents

\section{Introduction}

We can construct analogous fractional models of geometries and physical
theories in explicit form if we use fractional derivatives resulting in zero
for actions on constants (for instance, for the Caputo fractional
derivative). This is important for elaborating  geometric models of theories with
fractional calculus even (performing corresponding nonholonomic
deformations) we may prefer to work with another type of fractional
derivatives.

In this paper, we outline some key constructions for analogous classical and
quantum fractional theories \cite{vrfrf,vrfrg,bv1,bv2,bv3,bv4} when methods
of nonholonomic and Lagrange--Finsler geometry are generalized to fractional
dimensions.\footnote{%
we recommend readers to consult in advance the above cited papers on
details, notation conventions and bibliography}

An important consequence of such geometric approaches is that using
analogous and bi--Hamilton models (see integer dimension constructions \cite%
{vrflg,vacap,vanco}) and related solitonic systems we can study analytically
and numerically, as well to try to construct some analogous mechanical and
gravitational systems, with the aim to mimic a nonlinear/fractional
nonholonomic dynamics/evolution and even to provide certain schemes of
quantization, like in the ''fractional'' Fedosov approach \cite{bv2,vfed4}.

This work is organized in the form: In section \ref{s2}, we remember the
most important formulas on Caputo fractional derivatives and nonlinear
connections. Section \ref{s3} is devoted to fractional Lagrange--Finsler
geometries. There are presented the main constructions for analogous
fractional gravity in section \ref{s4}.

\vskip5pt \textbf{Acknowledgement: } This paper summarizes the results presented in our talk at the 3d Conference on ''Nonlinear Science and
Complexity'', 28--31 July, 2010, \c{C}hankaya University, Ankara, Turkey.

\section{Caputo Fractional Derivatives and N--connecti\-ons}

\label{s2}We summarize some important formulas on fractional calculus for
nonholonomic manifold elaborated in Refs. \cite{vrfrf,vrfrg,bv1,bv3}. Our
geometric arena consists from an abstract fractional manifold $\overset{%
\alpha }{\mathbf{V}}$ (we shall use also the term ''fractional space'' as an
equivalent one enabled with certain fundamental geometric structures) with
prescribed nonholonomic distribution modeling both the fractional calculus
and the non--integrable dynamics of interactions.

The fractional left, respectively, right Caputo derivatives are denoted in
the form
\begin{eqnarray}
&&\ _{\ _{1}x}\overset{\alpha }{\underline{\partial }}_{x}f(x):=\frac{1}{%
\Gamma (s-\alpha )}\int\limits_{\ \ _{1}x}^{x}(x-\ x^{\prime })^{s-\alpha
-1}\left( \frac{\partial }{\partial x^{\prime }}\right) ^{s}f(x^{\prime
})dx^{\prime };  \label{lfcd} \\
&&\ _{\ x}\overset{\alpha }{\underline{\partial }}_{\ _{2}x}f(x):=\frac{1}{%
\Gamma (s-\alpha )}\int\limits_{x}^{\ _{2}x}(x^{\prime }-x)^{s-\alpha
-1}\left( -\frac{\partial }{\partial x^{\prime }}\right) ^{s}f(x^{\prime
})dx^{\prime }\ .  \notag
\end{eqnarray}%
\ Using such operators, we can construct the fractional absolute
differential $\overset{\alpha }{d}:=(dx^{j})^{\alpha }\ \ _{\ 0}\overset{%
\alpha }{\underline{\partial }}_{j}$ when $\ \overset{\alpha }{d}%
x^{j}=(dx^{j})^{\alpha }\frac{(x^{j})^{1-\alpha }}{\Gamma (2-\alpha )},$
where we consider $\ _{1}x^{i}=0.$

We denote a fractional tangent bundle in the form $\overset{\alpha }{%
\underline{T}}M$ for $\alpha \in (0,1),$ associated to a manifold $M$ of
necessary smooth class and integer $\dim M=n.$\footnote{%
The symbol $T$ is underlined in order to emphasize that we shall associate
the approach to a fractional Caputo derivative.} Locally, both the integer
and fractional local coordinates are written in the form $u^{\beta
}=(x^{j},y^{a}).$ A fractional frame basis $\overset{\alpha }{\underline{e}}%
_{\beta }=e_{\ \beta }^{\beta ^{\prime }}(u^{\beta })\overset{\alpha }{%
\underline{\partial }}_{\beta ^{\prime }}$ on $\overset{\alpha }{\underline{T%
}}M$ \ is connected via a vierlbein transform $e_{\ \beta }^{\beta ^{\prime
}}(u^{\beta })$ with a fractional local coordinate basis
\begin{equation}
\overset{\alpha }{\underline{\partial }}_{\beta ^{\prime }}=\left( \overset{%
\alpha }{\underline{\partial }}_{j^{\prime }}=_{\ _{1}x^{j^{\prime }}}%
\overset{\alpha }{\underline{\partial }}_{j^{\prime }},\overset{\alpha }{%
\underline{\partial }}_{b^{\prime }}=_{\ _{1}y^{b^{\prime }}}\overset{\alpha
}{\underline{\partial }}_{b^{\prime }}\right) ,  \label{frlcb}
\end{equation}%
for $j^{\prime }=1,2,...,n$ and $b^{\prime }=n+1,n+2,...,n+n.$ The
fractional co--bases are written $\overset{\alpha }{\underline{e}}^{\ \beta
}=e_{\beta ^{\prime }\ }^{\ \beta }(u^{\beta })\overset{\alpha }{d}u^{\beta
^{\prime }},$ where the fractional local coordinate co--basis is
\begin{equation}
\ _{\ }\overset{\alpha }{d}u^{\beta ^{\prime }}=\left( (dx^{i^{\prime
}})^{\alpha },(dy^{a^{\prime }})^{\alpha }\right) .  \label{frlccb}
\end{equation}

It is possible to define a nonlinear connection (N--connection) $\overset{%
\alpha }{\mathbf{N}}$ \ for a fractional space $\overset{\alpha }{\mathbf{V}}
$ by a nonholonomic distribution (Whitney sum) with conventional h-- and
v--subspaces, $\underline{h}\overset{\alpha }{\mathbf{V}}$ and $\underline{v}%
\overset{\alpha }{\mathbf{V}},$%
\begin{equation}
\overset{\alpha }{\underline{T}}\overset{\alpha }{\mathbf{V}}=\underline{h}%
\overset{\alpha }{\mathbf{V}}\mathbf{\oplus }\underline{v}\overset{\alpha }{%
\mathbf{V}}.  \label{whit}
\end{equation}%
Locally, such a fractional N--connection is characterized by its local
coefficients $\overset{\alpha }{\mathbf{N}}\mathbf{=}\{\ ^{\alpha
}N_{i}^{a}\},$ when $\overset{\alpha }{\mathbf{N}}\mathbf{=}\ ^{\alpha
}N_{i}^{a}(u)(dx^{i})^{\alpha }\otimes \overset{\alpha }{\underline{\partial
}}_{a}.$

On $\overset{\alpha }{\mathbf{V}},$ it is convenient to work with N--adapted
fractional (co) frames,
\begin{eqnarray}
\ ^{\alpha }\mathbf{e}_{\beta } &=&\left[ \ ^{\alpha }\mathbf{e}_{j}=\overset%
{\alpha }{\underline{\partial }}_{j}-\ ^{\alpha }N_{j}^{a}\overset{\alpha }{%
\underline{\partial }}_{a},\ ^{\alpha }e_{b}=\overset{\alpha }{\underline{%
\partial }}_{b}\right] ,  \label{dder} \\
\ ^{\alpha }\mathbf{e}^{\beta } &=&[\ ^{\alpha }e^{j}=(dx^{j})^{\alpha },\
^{\alpha }\mathbf{e}^{b}=(dy^{b})^{\alpha }+\ ^{\alpha
}N_{k}^{b}(dx^{k})^{\alpha }].  \label{ddif}
\end{eqnarray}

A fractional metric structure (d--metric) $\ \overset{\alpha }{\mathbf{g}}%
=\{\ ^{\alpha }g_{\underline{\alpha }\underline{\beta }}\}=\left[ \ ^{\alpha
}g_{kj},\ ^{\alpha }g_{cb}\right] $ on $\overset{\alpha }{\mathbf{V}}$ \ can
be represented in different equivalent forms,
\begin{eqnarray}
\ \overset{\alpha }{\mathbf{g}} &=&\ ^{\alpha }g_{\underline{\gamma }%
\underline{\beta }}(u)(du^{\underline{\gamma }})^{\alpha }\otimes (du^{%
\underline{\beta }})^{\alpha }  \label{m1} \\
&=&\ ^{\alpha }g_{kj}(x,y)\ ^{\alpha }e^{k}\otimes \ ^{\alpha }e^{j}+\
^{\alpha }g_{cb}(x,y)\ ^{\alpha }\mathbf{e}^{c}\otimes \ ^{\alpha }\mathbf{e}%
^{b}  \notag \\
&=&\eta _{k^{\prime }j^{\prime }}\ ^{\alpha }e^{k^{\prime }}\otimes \
^{\alpha }e^{j^{\prime }}+\eta _{c^{\prime }b^{\prime }}\ ^{\alpha }\mathbf{e%
}^{c^{\prime }}\otimes \ ^{\alpha }\mathbf{e}^{b^{\prime }},  \notag
\end{eqnarray}%
where matrices $\eta _{k^{\prime }j^{\prime }}=diag[\pm 1,\pm 1,...,\pm 1]$
and $\eta _{a^{\prime }b^{\prime }}=diag[\pm 1,\pm 1,...,\pm 1],$ for the
signature of a ''prime'' spacetime $\mathbf{V,}$ are obtained by frame
transforms $\eta _{k^{\prime }j^{\prime }}=e_{\ k^{\prime }}^{k}\ e_{\
j^{\prime }}^{j}\ _{\ }^{\alpha }g_{kj}$ and $\eta _{a^{\prime }b^{\prime
}}=e_{\ a^{\prime }}^{a}\ e_{\ b^{\prime }}^{b}\ _{\ }^{\alpha }g_{ab}.$

We can adapt geometric objects on $\overset{\alpha }{\mathbf{V}}$ with
respect to a given N--connection structure $\overset{\alpha }{\mathbf{N}},$
calling them as distinguished objects (d--objects). For instance, a
distinguished connection (d--connection) $\overset{\alpha }{\mathbf{D}}$ on $%
\overset{\alpha }{\mathbf{V}}$ is defined as a linear connection preserving
under parallel transports the Whitney sum (\ref{whit}). There is an
associated N--adapted differential 1--form
\begin{equation}
\ ^{\alpha }\mathbf{\Gamma }_{\ \beta }^{\tau }=\ ^{\alpha }\mathbf{\Gamma }%
_{\ \beta \gamma }^{\tau }\ ^{\alpha }\mathbf{e}^{\gamma },  \label{fdcf}
\end{equation}%
parametrizing the coefficients (with respect to (\ref{ddif}) and (\ref{dder}%
)) in the form $\ ^{\alpha }\mathbf{\Gamma }_{\ \tau \beta }^{\gamma
}=\left( \ ^{\alpha }L_{jk}^{i},\ ^{\alpha }L_{bk}^{a},\ ^{\alpha
}C_{jc}^{i},\ ^{\alpha }C_{bc}^{a}\right) .$

The absolute fractional differential $\ ^{\alpha }\mathbf{d}=\ _{\ _{1}x}%
\overset{\alpha }{d}_{x}+\ _{\ _{1}y}\overset{\alpha }{d}_{y}$ acts on
fractional differential forms in N--adapted form. This is a fractional
distinguished operator, d--operator, when the value $\ ^{\alpha }\mathbf{d:=}%
\ ^{\alpha }\mathbf{e}^{\beta }\ ^{\alpha }\mathbf{e}_{\beta }$ splits into
exterior h- and v--derivatives when
\begin{equation*}
\ _{\ _{1}x}\overset{\alpha }{d}_{x}:=(dx^{i})^{\alpha }\ \ _{\ _{1}x}%
\overset{\alpha }{\underline{\partial }}_{i}=\ ^{\alpha }e^{j}\ ^{\alpha }%
\mathbf{e}_{j}\mbox{ and }_{\ _{1}y}\overset{\alpha }{d}_{y}:=(dy^{a})^{%
\alpha }\ \ _{\ _{1}x}\overset{\alpha }{\underline{\partial }}_{a}=\
^{\alpha }\mathbf{e}^{b}\ ^{\alpha }e_{b}.
\end{equation*}%
Using such differentials, we can compute in explicit form the torsion and
curvature (as fractional two d--forms derived for (\ref{fdcf})) of a
fractional d--connection $\overset{\alpha }{\mathbf{D}}=\{\ ^{\alpha }%
\mathbf{\Gamma }_{\ \beta \gamma }^{\tau }\},$
\begin{eqnarray}
\ ^{\alpha }\mathcal{T}^{\tau } &\doteqdot &\overset{\alpha }{\mathbf{D}}\
^{\alpha }\mathbf{e}^{\tau }=\ ^{\alpha }\mathbf{d}\ ^{\alpha }\mathbf{e}%
^{\tau }+\ ^{\alpha }\mathbf{\Gamma }_{\ \beta }^{\tau }\wedge \ ^{\alpha }%
\mathbf{e}^{\beta }\mbox{ and }  \label{tors} \\
\ ^{\alpha }\mathcal{R}_{~\beta }^{\tau } &\doteqdot &\overset{\alpha }{%
\mathbf{D}}\mathbf{\ ^{\alpha }\Gamma }_{\ \beta }^{\tau }=\ ^{\alpha }%
\mathbf{d\ ^{\alpha }\Gamma }_{\ \beta }^{\tau }-\ ^{\alpha }\mathbf{\Gamma }%
_{\ \beta }^{\gamma }\wedge \ ^{\alpha }\mathbf{\Gamma }_{\ \gamma }^{\tau
}=\ ^{\alpha }\mathbf{R}_{\ \beta \gamma \delta }^{\tau }\ ^{\alpha }\mathbf{%
e}^{\gamma }\wedge \ ^{\alpha }\mathbf{e}^{\delta }.  \notag
\end{eqnarray}

Contracting respectively the indices, we can compute the fractional Ricci
tensor $\ ^{\alpha }\mathcal{R}ic=\{\ ^{\alpha }\mathbf{R}_{\alpha \beta
}\doteqdot \ ^{\alpha }\mathbf{R}_{\ \alpha \beta \tau }^{\tau }\}$ with
components
\begin{equation}
\ ^{\alpha }R_{ij}\doteqdot \ ^{\alpha }R_{\ ijk}^{k},\ \ \ ^{\alpha
}R_{ia}\doteqdot -\ ^{\alpha }R_{\ ika}^{k},\ \ ^{\alpha }R_{ai}\doteqdot \
^{\alpha }R_{\ aib}^{b},\ \ ^{\alpha }R_{ab}\doteqdot \ ^{\alpha }R_{\
abc}^{c}  \label{dricci}
\end{equation}%
and the scalar curvature of fractional d--connection $\overset{\alpha }{%
\mathbf{D}},$
\begin{equation}
\ _{s}^{\alpha }\mathbf{R}\doteqdot \ ^{\alpha }\mathbf{g}^{\tau \beta }\
^{\alpha }\mathbf{R}_{\tau \beta }=\ ^{\alpha }R+\ ^{\alpha }S,\ ^{\alpha
}R=\ ^{\alpha }g^{ij}\ ^{\alpha }R_{ij},\ \ ^{\alpha }S=\ ^{\alpha }g^{ab}\
^{\alpha }R_{ab},  \label{sdccurv}
\end{equation}%
with $\ ^{\alpha }\mathbf{g}^{\tau \beta }$ being the inverse coefficients
to a d--metric (\ref{m1}).

The Einstein tensor of any metric compatible $\overset{\alpha }{\mathbf{D}},
$ when $\overset{\alpha }{\mathbf{D}}_{\tau }\ ^{\alpha }\mathbf{g}^{\tau
\beta }=0,$ is defined $\ ^{\alpha }\mathcal{E}ns=\{\ ^{\alpha }\mathbf{G}%
_{\alpha \beta }\},$ where
\begin{equation}
\ ^{\alpha }\mathbf{G}_{\alpha \beta }:=\ ^{\alpha }\mathbf{R}_{\alpha \beta
}-\frac{1}{2}\ ^{\alpha }\mathbf{g}_{\alpha \beta }\ \ _{s}^{\alpha }\mathbf{%
R.}  \label{enstdt}
\end{equation}

The regular fractional mechanics defined by a fractional Lagrangian $\overset%
{\alpha }{L}$ can be equivalently encoded into canonical geometric data $(\
_{L}\overset{\alpha }{\mathbf{N}},\ _{L}\overset{\alpha }{\mathbf{g}},\
_{c}^{\alpha }\mathbf{D}),$ where we put the label $L$ in order to emphasize
that such geometric objects are induced by a fractional Lagrangian as we
provided in \cite{vrfrf,vrfrg,bv1,bv3}. We also note that it is possible to
''arrange'' on $\overset{\alpha }{\mathbf{V}}$ such nonholonomic
distributions when a d--connection $\ \ _{0}\overset{\alpha }{\mathbf{D}}%
=\{\ _{0}^{\alpha }\widetilde{\mathbf{\Gamma }}_{\ \alpha ^{\prime }\beta
^{\prime }}^{\gamma ^{\prime }}\}$ is described by constant matrix
coefficients, see details in \cite{vacap,vanco}, for integer dimensions, and %
\cite{bv3}, for fractional dimensions.

\section{Fractional Lagrange--Finsler Geometry}

\label{s3} A Lagrange space $L^{n}=(M,L),$ of integer dimension $n,$ is
defined by a Lagrange fundamental function $L(x,y),$ i.e. a regular real function $L:$ $TM\rightarrow \mathbb{R},$ for which the Hessian
$\ _{L}g_{ij}=(1/2)\partial ^{2}L/\partial y^{i}\partial y^{j}$
 is not degenerate.

We say that a Lagrange space $L^{n}$ is a Finsler space $F^{n}$ if and only
if its fundamental function $L$ is positive and two homogeneous with respect
to variables $y^{i},$ i.e. $L=F^{2}.$ For simplicity, we shall work with
Lagrange spaces and their fractional generalizations, considering the
Finsler ones to consist of a more particular, homogeneous, subclass.

\begin{definition}
A (target) fractional Lagrange space $\overset{\alpha }{\underline{L^{n}}}=(%
\overset{\alpha }{\underline{M}},\overset{\alpha }{L})$ of \ fractional
dimension $\alpha \in (0,1),$ for a regular real function $\overset{\alpha }{%
L}:$ $\overset{\alpha }{\underline{T}}M\rightarrow \mathbb{R},$ when the
fractional Hessian is
\begin{equation}
\ _{L\ }\overset{\alpha }{g}_{ij}=\frac{1}{4}\left( \overset{\alpha }{%
\underline{\partial }}_{i}\overset{\alpha }{\underline{\partial }}_{j}+%
\overset{\alpha }{\underline{\partial }}_{j}\overset{\alpha }{\underline{%
\partial }}_{i}\right) \overset{\alpha }{L}\neq 0.  \label{hessf}
\end{equation}
\end{definition}

In our further constructions, we shall use the coefficients $\ _{L\ }\overset%
{\alpha }{g^{ij}\text{ }}$being inverse to $_{L\ }\overset{\alpha }{g}_{ij}$
(\ref{hessf}).\footnote{%
We shall put a left label $L$ to certain geometric objects if it is
necessary to emphasize that they are induced by Lagrange generating
function. Nevertheless, such labels will be omitted (in order to simplify
the notations) if that will not result in ambiguities.} Any $\overset{\alpha
}{\underline{L^{n}}}$ can be associated to a prime ''integer'' Lagrange
space $L^{n}.$

The concept of nonlinear connection (N--connection) on $\overset{\alpha }{%
\underline{L^{n}}}$ can be introduced similarly to that on nonholonomic
fractional manifold \cite{vrfrf,vrfrg} considering the fractional tangent
bundle $\overset{\alpha }{\underline{T}}M.$

\begin{definition}
A N--connection $\overset{\alpha }{\mathbf{N}}$ on $\overset{\alpha }{%
\underline{T}}M$ is defined by a nonholonomic distribution (Whitney sum)
with conventional h-- and v--subspaces, $\underline{h}$ $\overset{\alpha }{%
\underline{T}}M$ and $\underline{v}\overset{\alpha }{\underline{T}}M,$ when
\begin{equation}
\ \overset{\alpha }{\underline{T}}\overset{\alpha }{\underline{T}}M=%
\underline{h}\overset{\alpha }{\underline{T}}M\mathbf{\oplus }\underline{v}%
\overset{\alpha }{\underline{T}}M.  \label{whitney}
\end{equation}
\end{definition}

Locally, a fractional N--connection is defined by a set of coefficients, $%
\overset{\alpha }{\mathbf{N}}\mathbf{=}\{\ ^{\alpha }N_{i}^{a}\},$ when
\begin{equation}
\overset{\alpha }{\mathbf{N}}\mathbf{=}\ \ ^{\alpha
}N_{i}^{a}(u)(dx^{i})^{\alpha }\otimes \overset{\alpha }{\underline{\partial
}}_{a},  \label{fnccoef}
\end{equation}%
see local bases (\ref{frlcb}) and (\ref{frlccb}).

Let$\ $\ us consider values $y^{k}(\tau )=dx^{k}(\tau )/d\tau ,$ for $x(\tau
)$ parametrizing smooth curves on a manifold $M$ \ with $\tau \in \lbrack
0,1].$ The fractional analogs of such configurations are determined by
changing $\ d/d\tau $ into the fractional Caputo derivative $\ \overset{%
\alpha }{\underline{\partial }}_{\tau }=_{\ _{1}\tau }\overset{\alpha }{%
\underline{\partial }}_{\tau }$when $\ ^{\alpha }y^{k}(\tau )=\overset{%
\alpha }{\underline{\partial }}_{\tau }x^{k}(\tau ).$ For simplicity, we
shall omit the label $\alpha $ for $y\in $ $\overset{\alpha }{\underline{T}}M
$ if that will not result in ambiguities and/or we shall do not associate to
it an explicit fractional derivative along a curve.

By straightforward computations, following the same scheme as in \cite%
{vrflg} but with fractional derivatives and integrals, we prove:

\begin{theorem}
Any $\overset{\alpha }{L}$ defines the fundamental geometric objects
determining canonically a nonholonomic fractional Riemann--Cartan geometry
on $\overset{\alpha }{\underline{T}}M$ being satisfied the properties:

\begin{enumerate}
\item The fractional Euler--Lagrange equations%
\begin{equation*}
\ \overset{\alpha }{\underline{\partial }}_{\tau \ }(\ _{\ _{1}y^{i}}\overset%
{\alpha }{\underline{\partial }}_{i}\overset{\alpha }{L})-_{\ _{1}x^{i}}%
\overset{\alpha }{\underline{\partial }}_{i}\overset{\alpha }{L}=0
\end{equation*}
are equivalent to the fractional ''nonlinear geodesic'' (equivalently,
semi--spray) equations $\ $%
\begin{equation*}
\left( \overset{\alpha }{\underline{\partial }}_{\tau \ }\right) ^{2}x^{k}+2%
\overset{\alpha }{G^{k}}(x,\ ^{\alpha }y)=0,
\end{equation*}
where
\begin{equation*}
\overset{\alpha }{G^{k}}=\frac{1}{4}\ \ _{L\ }\overset{\alpha }{g^{kj}}\left[
y^{j}\ _{\ _{1}y^{j}}\overset{\alpha }{\underline{\partial }}_{j}\ \left(
_{\ _{1}x^{i}}\overset{\alpha }{\underline{\partial }}_{i}\overset{\alpha }{L%
}\right) -\ _{\ _{1}x^{i}}\overset{\alpha }{\underline{\partial }}_{i}%
\overset{\alpha }{L}\right]
\end{equation*}%
defines the canonical N--connection $\ $%
\begin{equation}
\ _{L}^{\alpha }N_{j}^{a}=\ _{\ _{1}y^{j}}\overset{\alpha }{\underline{%
\partial }}_{j}\overset{\alpha }{G^{k}}(x,\ ^{\alpha }y).  \label{cncl}
\end{equation}

\item There is a canonical (Sasaki type) metric structure,
\begin{equation*}
\ \ _{L}\overset{\alpha }{\mathbf{g}}=\ _{L}^{\alpha }g_{kj}(x,y)\ ^{\alpha
}e^{k}\otimes \ ^{\alpha }e^{j}+\ _{L}^{\alpha }g_{cb}(x,y)\ _{L}^{\alpha }%
\mathbf{e}^{c}\otimes \ _{L}^{\alpha }\mathbf{e}^{b},  \label{sasm}
\end{equation*}%
where the preferred frame structure (defined linearly by $\ \ _{L}^{\alpha
}N_{j}^{a})$ is $\ _{L}^{\alpha }\mathbf{e}_{\nu }=(\ _{L}^{\alpha }\mathbf{e%
}_{i},e_{a}).$

\item There is a canonical metrical distinguished connection
\begin{equation*}
\ _{c}^{\alpha }\mathbf{D}=(h\ _{c}^{\alpha }D,v\ _{c}^{\alpha }D)=\{\
_{c}^{\alpha }\mathbf{\Gamma }_{\ \alpha \beta }^{\gamma }=(\ ^{\alpha }%
\widehat{L}_{\ jk}^{i},\ ^{\alpha }\widehat{C}_{jc}^{i})\}\ ,
\end{equation*}%
(in brief, d--connection), which is a linear connection preserving under
parallelism the splitting (\ref{whitney}) and metric compatible, i.e. $\
_{c}^{\alpha }\mathbf{D}\ \left( \ \ _{L}\overset{\alpha }{\mathbf{g}}%
\right) =0,$ when
\begin{equation*}
\ _{c}^{\alpha }\mathbf{\Gamma }_{\ j}^{i}=\ _{c}^{\alpha }\mathbf{\Gamma }%
_{\ j\gamma }^{i}\ _{L}^{\alpha }\mathbf{e}^{\gamma }=\widehat{L}_{\
jk}^{i}e^{k}+\widehat{C}_{jc}^{i}\ _{L}^{\alpha }\mathbf{e}^{c},
\end{equation*}%
for $\widehat{L}_{\ jk}^{i}=\widehat{L}_{\ bk}^{a},\widehat{C}_{jc}^{i}=%
\widehat{C}_{bc}^{a}$ in $\ \ _{c}^{\alpha }\mathbf{\Gamma }_{\ b}^{a}=\
_{c}^{\alpha }\mathbf{\Gamma }_{\ b\gamma }^{a}\ _{L}^{\alpha }\mathbf{e}%
^{\gamma }=\widehat{L}_{\ bk}^{a}e^{k}+\widehat{C}_{bc}^{a}\ _{L}^{\alpha }%
\mathbf{e}^{c},$
\begin{eqnarray*}
\ ^{\alpha }\widehat{L}_{jk}^{i} &=&\frac{1}{2}\ _{L}^{\alpha }g^{ir}\left(
\ _{L}^{\alpha }\mathbf{e}_{k}\ _{L}^{\alpha }g_{jr}+\ _{L}^{\alpha }\mathbf{%
e}_{j}\ _{L}^{\alpha }g_{kr}-\ _{L}^{\alpha }\mathbf{e}_{r}\ _{L}^{\alpha
}g_{jk}\right) ,  \label{cdc} \\
\ \ ^{\alpha }\widehat{C}_{bc}^{a} &=&\frac{1}{2}\ _{L}^{\alpha
}g^{ad}\left( \ ^{\alpha }e_{c}\ _{L}^{\alpha }g_{bd}+\ ^{\alpha }e_{c}\
_{L}^{\alpha }g_{cd}-\ ^{\alpha }e_{d}\ _{L}^{\alpha }g_{bc}\right)  \notag
\end{eqnarray*}%
are just the generalized Christoffel indices.\footnote{%
for integer dimensions, we contract ''horizontal'' and ''vertical'' indices
following the rule: $i=1$ is $a=n+1;$ $i=2$ is $a=n+2;$ ... $i=n$ is $a=n+n"$%
}
\end{enumerate}
\end{theorem}

Finally, in this section, we note that:

\begin{remark}
We note that $\ _{c}^{\alpha }\mathbf{D}$ is with nonholonomically induced
torsion structure defined by 2--forms%
\begin{eqnarray*}
\ _{L}^{\alpha }\mathcal{T}^{i} &=&\widehat{C}_{\ jc}^{i}\ ^{\alpha
}e^{i}\wedge \ _{L}^{\alpha }\mathbf{e}^{c},  \label{nztors} \\
\ _{L}^{\alpha }\mathcal{T}^{a} &=&-\frac{1}{2}\ _{L}\Omega _{ij}^{a}\
^{\alpha }e^{i}\wedge \ ^{\alpha }e^{j}+\left( \ ^{\alpha }e_{b}\
_{L}^{\alpha }N_{i}^{a}-\ ^{\alpha }\widehat{L}_{\ bi}^{a}\right) \ ^{\alpha
}e^{i}\wedge \ _{L}^{\alpha }\mathbf{e}^{b}  \notag
\end{eqnarray*}%
computed from the fractional version of Cartan's structure equations%
\begin{eqnarray*}
d\ ^{\alpha }e^{i}-\ ^{\alpha }e^{k}\wedge \ \ _{c}^{\alpha }\mathbf{\Gamma }%
_{\ k}^{i} &=&-\ _{L}^{\alpha }\mathcal{T}^{i},\   \notag \\
d\ _{L}^{\alpha }\mathbf{e}^{a}-\ _{L}^{\alpha }\mathbf{e}^{b}\wedge \ \
_{c}^{\alpha }\mathbf{\Gamma }_{\ b}^{a} &=&-\ _{L}^{\alpha }\mathcal{T}^{a},
\notag \\
d\ \ _{c}^{\alpha }\mathbf{\Gamma }_{\ j}^{i}-\ \ _{c}^{\alpha }\mathbf{%
\Gamma }_{\ j}^{k}\wedge \ \ _{c}^{\alpha }\mathbf{\Gamma }_{\ k}^{i} &=&-\
_{L}^{\alpha }\mathcal{R}_{j}^{i}  \label{seq}
\end{eqnarray*}%
in which the curvature 2--form is denoted $\ _{L}^{\alpha }\mathcal{R}%
_{j}^{i}.$
\end{remark}

In general, for any d--connection on $\ \overset{\alpha }{\underline{T}}M,$
we can compute respectively the N--adapted coefficients of torsion $\
^{\alpha }\mathcal{T}^{\tau }=\{\ ^{\alpha }\mathbf{\Gamma }_{\ \beta \gamma
}^{\tau }\}$ and curvature $\ ^{\alpha }\mathcal{R}_{~\beta }^{\tau }=\{\
^{\alpha }\mathbf{R}_{\ \beta \gamma \delta }^{\tau }\}$ as it is explained
for general fractional nonholonomic manifolds in \cite{vrfrf,vrfrg}.

\section{Analogous Fractional Gravity}

\label{s4}Let us consider a ''prime'' nonholonomic manifold $\mathbf{V}$ is
of integer dimension $\dim $ $\mathbf{V}=n+m,n\geq 2,m\geq 1.$\footnote{%
A nonholonomic manifold is a manifold endowed with a non--integrable
(equivalently, nonholonomic, or anholonomic) distribution. There are three
useful (for our considerations) examples when 1) $\mathbf{V}$ is a (pseudo)
Riemannian manifold; 2) $\mathbf{V}=E(M),$ or 3)\ $\mathbf{V}=TM,$ for a
vector, or tangent, bundle on a base manifold $M.$ We also emphasize that in
this paper we follow the conventions from Refs. \cite{vrflg,vrfrf,vrfrg}
when left indices are used as labels and right indices may be abstract ones
or running certain values.}\ Its fractional extension $\overset{\alpha }{%
\mathbf{V}}$ is modelled by a quadruple $(\mathbf{V},\overset{\alpha }{%
\mathbf{N}},\overset{\alpha }{\mathbf{d}},\overset{\alpha }{\mathbf{I}}),$
where $\overset{\alpha }{\mathbf{N}}$ is a nonholonomic distribution stating
a nonlinear connection (N--connection) structure. The fractional
differential structure $\overset{\alpha }{\mathbf{d}}$ is determined by
Caputo fractional derivative (\ref{lfcd}) following formulas (\ref{frlcb})
and (\ref{frlccb}).

For any respective frame and co--frame (dual) structures,\newline
$\ ^{\alpha }e_{\alpha ^{\prime }}=(\ ^{\alpha }e_{i^{\prime }},\ ^{\alpha
}e_{a^{\prime }})$ and $\ ^{\alpha }e_{\ }^{\beta ^{\prime }}=(\ ^{\alpha
}e^{i^{\prime }},\ ^{\alpha }e^{a^{\prime }})$ \ on $\overset{\alpha }{%
\mathbf{V}}\mathbf{,}$ we can consider \ frame transforms
\begin{equation}
\ ^{\alpha }e_{\alpha }=A_{\alpha }^{\ \alpha ^{\prime }}(x,y)\ ^{\alpha
}e_{\alpha ^{\prime }}\mbox{\ and\ }\ ^{\alpha }e_{\ }^{\beta }=A_{\ \beta
^{\prime }}^{\beta }(x,y)\ ^{\alpha }e^{\beta ^{\prime }}.  \label{nhft}
\end{equation}

A subclass of frame transforms (\ref{nhft}), for fixed ''prime'' and
''target'' frame structures, is called N--adapted if such nonholonomic
transformations preserve the splitting defined by a N--connection structure $%
\mathbf{N}=\{N_{i}^{a}\}.$

Under (in general, nonholonomic) frame transforms, the metric coefficients
of any metric structure $\overset{\alpha }{\mathbf{g}}$ on $\overset{\alpha }%
{\mathbf{V}}$ are re--computed following formulas
\begin{equation*}
\ ^{\alpha }g_{\alpha \beta }(x,y)=A_{\alpha }^{~\ \alpha ^{\prime
}}(x,y)~A_{\beta }^{\ \beta ^{\prime }}(x,y)\ ^{\alpha }g_{\alpha ^{\prime
}\beta ^{\prime }}(x,y).
\end{equation*}

For any fixed \ $\overset{\alpha }{\mathbf{g}}$ and $\overset{\alpha }{%
\mathbf{N}},$ there are N--adapted frame transforms when
\begin{eqnarray*}
\overset{\alpha }{\mathbf{g}} &=&\ ^{\alpha }g_{ij}(x,y)\ \ ^{\alpha
}e^{i}\otimes \ ^{\alpha }e^{j}+\ ^{\alpha }h_{ab}(x,y)\ ^{\alpha }\mathbf{e}%
^{a}\otimes \ ^{\alpha }\mathbf{e}^{b}, \\
&=&\ ^{\alpha }g_{i^{\prime }j^{\prime }}(x,y)\ \ ^{\alpha }e^{i^{\prime
}}\otimes \ ^{\alpha }e^{j^{\prime }}+\ ^{\alpha }h_{a^{\prime }b^{\prime
}}(x,y)\ \ ^{\alpha }\mathbf{e}^{a^{\prime }}\otimes \ ^{\alpha }\mathbf{e}%
^{b^{\prime }},
\end{eqnarray*}%
where $\ ^{\alpha }\mathbf{e}^{a}$ and $\ ^{\alpha }\mathbf{e}^{a^{\prime }}$
are elongated following formulas (\ref{ddif}), respectively by $\ ^{\alpha
}N_{\ j}^{a}$ and
\begin{equation}
\ ^{\alpha }N_{\ j^{\prime }}^{a^{\prime }}=A_{a}^{~\ a^{\prime }}(x,y)A_{\
j^{\prime }}^{j}(x,y)\ ^{\alpha }N_{\ j}^{a}(x,y),  \label{ncontri}
\end{equation}%
or, inversely,
\begin{equation*}
\ ^{\alpha }N_{\ j}^{a}=A_{a^{\prime }}^{~\ a}(x,y)A_{\ j}^{j^{\prime
}}(x,y)\ ^{\alpha }N_{\ j^{\prime }}^{a^{\prime }}(x,y)
\end{equation*}%
with prescribed $\ ^{\alpha }N_{\ j^{\prime }}^{a^{\prime }}.$

We preserve the N--connection splitting for any frame transform of type (\ref%
{nhft}) when
\begin{equation*}
\ ^{\alpha }g_{i^{\prime }j^{\prime }}=A_{\ i^{\prime }}^{i}A_{\ j^{\prime
}}^{j}\ ^{\alpha }g_{ij},\ \ ^{\alpha }h_{a^{\prime }b^{\prime }}=A_{\
a^{\prime }}^{a}A_{\ b^{\prime }}^{b}\ ^{\alpha }h_{ab},
\end{equation*}%
for $A_{i}^{~\ i^{\prime }}$ constrained to get holonomic $\ ^{\alpha
}e^{i^{\prime }}=A_{i}^{~\ i^{\prime }}\ ^{\alpha }e^{i},$ i.e. $[\ ^{\alpha
}e^{i^{\prime }},\ ^{\alpha }e^{j^{\prime }}]=0$ and $\ ^{\alpha }\mathbf{e}%
^{a^{\prime }}=dy^{a^{\prime }}+\ ^{\alpha }N_{\ j^{\prime }}^{a^{\prime
}}dx^{j^{\prime }},$ for certain $x^{i^{\prime }}=x^{i^{\prime
}}(x^{i},y^{a})$ and $y^{a^{\prime }}=y^{a^{\prime }}(x^{i},y^{a}),$ with $\
^{\alpha }N_{\ j^{\prime }}^{a^{\prime }}$ computed following formulas (\ref%
{ncontri}). Such conditions can be satisfied by prescribing from the very
beginning a nonholonomic distribution of necessary type. The constructions
can be equivalently inverted, when $\ ^{\alpha }g_{\alpha \beta }$ and $\
^{\alpha }N_{i}^{a}$ are computed from $\ ^{\alpha }g_{\alpha ^{\prime
}\beta ^{\prime }}$ and $\ ^{\alpha }N_{i^{\prime }}^{a^{\prime }},$ if both
the metric and N--connection splitting structures are fixed on $\overset{%
\alpha }{\mathbf{V}}.$

An unified approach to Einstein--Lagrange/Finsler gravity for arbitrary
integer and non--integer dimensions is possible for the fractional canonical
d--connection $\ ^{\alpha }\widehat{\mathbf{D}}.$ The fractional
gravitational field equations are formulated for the Einstein d--tensor (\ref%
{enstdt}), following the same principle of constructing the matter source $\
^{\alpha }\mathbf{\Upsilon }_{\beta \delta }$ as in general relativity but
for fractional metrics and d--connections,%
\begin{equation*}
\ ^{\alpha }\widehat{\mathbf{E}}_{\ \beta \delta }=\ ^{\alpha }\mathbf{%
\Upsilon }_{\beta \delta }.  \label{fdeq}
\end{equation*}%
Such a system of integro--differential equations for generalized connections
can be restricted to fractional nonholonomic configurations for $\ ^{\alpha
}\nabla $ if we impose the additional constraints%
\begin{equation*}
\ ^{\alpha }\widehat{L}_{aj}^{c}=\ ^{\alpha }e_{a}(\ ^{\alpha }N_{j}^{c}),\
\ ^{\alpha }\widehat{C}_{jb}^{i}=0,\ \ ^{\alpha }\Omega _{\ ji}^{a}=0.
\label{frconstr}
\end{equation*}

There are not theoretical or experimental evidences that for fractional
dimensions we must impose conditions of type (\ref{frconstr}) but they have
certain physical motivation if we develop models which in integer limits
result in the general relativity theory.


\begin{thebibliography}{99}
\bibitem{vrfrf} S. Vacaru, Fractional Nonholonomic Ricci Flows, arXiv:
1004.0625

\bibitem{vrfrg} S. Vacaru, Fractional Dynamics from Einstein Gravity,
General Solutions, and Black Holes, arXiv: 1004.0628

\bibitem{bv1} D. Baleanu and S. Vacaru, Fractional Almost Kahler-Lagrange
Geometry, arXiv: 1006.5535

\bibitem{bv2} D. Baleanu and S. Vacaru, Fedosov Quantization of Fractional
Lagrange Spaces, arXiv: 1006.5538

\bibitem{bv3} D. Baleanu and S. Vacaru, Constant Curvature Coefficients and
Exact Solutions in Fractional Gravity and Geometric Mechanics, arXiv:
1007.2864

\bibitem{bv4} D. Baleanu and S. Vacaru, Fractional Curve Flows and Solitonic
Hierarchies in Gravity and Geometric Mechanics, arXiv: 1007.2866

\bibitem{vrflg} S. Vacaru, Finsler and Lagrange geometries in Einstein and
string gravity, Int. J. Geom. Methods. Mod. Phys. (IJGMMP) \textbf{5} (2008)
473-51

\bibitem{vfed4} S. Vacaru, Einstein Gravity as a Nonholonomic Almost Kahler
Geometry, Lagrange-Finsler Variables, and Deformation Quantization, J. Geom.
Phys. \textbf{\ 60 } (2010) 1289-13051

\bibitem{vacap} S. Vacaru, Curve Flows and Solitonic Hierarchies Generated
by Einstein Metrics, Acta Applicandae Mathematicae [ACAP] \textbf{110}
(2010) 73-107; arXiv: 0810.0707

\bibitem{vanco} S. Anco and S. Vacaru, Curve Flows in Lagrange-Finsler
Geometry, Bi-Hamiltonian Structures and Solitons, J. Geom. Phys. \textbf{59}
(2009) 79-103
\end{thebibliography}
\end{document}